\documentclass[aps,prc,groupedaddress,twocolumn,showpacs,amsmath,amssymb,floatfix]{revtex4}
\usepackage{graphics}
\usepackage{dcolumn}
\usepackage{longtable}

\begin{document}

\title{Charged--particle optical potentials tested by first direct measurement of $^{59}$Cu$(p,\alpha)^{56}$Ni reaction}
\author{V.~Avrigeanu} \email{vlad.avrigeanu@nipne.ro}
\author{M.~Avrigeanu}
\affiliation{Horia Hulubei National Institute for Physics and Nuclear Engineering, 077125 Bucharest-Magurele, Romania}

\begin{abstract}
Due consideration of proton optical--model potential (OMP) anomalies at sub-Coulomb energies for medium--weight nuclei is shown to be critical for the analysis of the unprecedented measurement of $^{59}$Cu$(p,\alpha)^{56}$Ni reaction cross section at an energy of $\sim$6 MeV [Phys. Rev. C {\bf 104}, L042801 (2021)]. 
The variation in predicted cross sections from standard statistical--model calculations and the cross--section range corresponding to the anomalous proton imaginary--potential depth, for target nuclei off the line of stability, are distinct and well separated.
Consequently, the new measurement provides, under unique conditions, tests of proton isoscalar and isovector real--potential components, the anomalous imaginary potential, as well as previous alpha-particle OMP, for nuclei off the line of stability.
\pacs{24.10.Ht,24.60.Dr,25.40.-h,27.50.+e}
\end{abstract}

\maketitle


Following a first direct measurement of $^{59}$Cu$(p,\alpha)^{56}$Ni reaction cross section at a center-of-mass energy of 6 MeV \cite{jsr21}, a reaction modeling challenge becomes possible on far better terms than ever before. 
This reaction $Q$-value of +2.413 MeV and the first excited state of the residual double--magic nucleus $^{56}$Ni at 2.701 MeV led at this energy to a real competition of merely inelastic scattering and $(p,\alpha)$ reaction to $^{56}$Ni ground state. 
In such a case, calculated cross sections within Hauser--Feshbach (HF) statistical model were assumed essentially sensitive only to the $\alpha$-particle optical model potential (OMP) whereas other ingredients like the nucleon OMP, the $\gamma$-ray strength function, and the level density have only marginal influence \cite{jsr21}. 
However, it was found that all recent $\alpha$-particle OMPs, including that of \cite{va14}, overestimate the new experimental result by a factor of 2. 

On the other hand, in an enlarged analysis of nucleon--induced alpha emission in the mass range $A$$\sim$60, a suitable account of $(p,\alpha)$ reaction on $^{63,65}$Cu stable isotopes has been found at similar incident energies \cite{va21}. 
Moreover, it has also involved the $\alpha$-particle OMP \cite{va14}, but with no overestimation as the above--mentioned. 
%
Therefore, we have found of interest a similar analysis for $^{59}$Cu$(p,\alpha)^{56}$Ni reaction cross section also related to a distinct nucleus off the line of stability. 

\begin{figure} 
\resizebox{1.0\columnwidth}{!}{\includegraphics{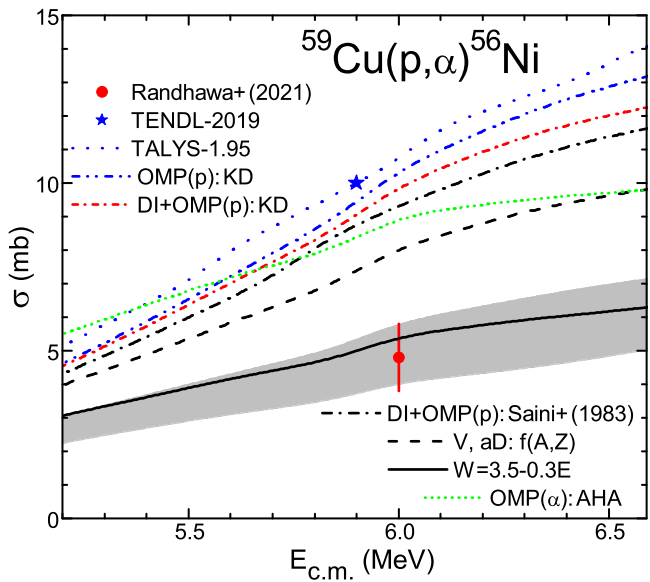}}
\caption{\label{Fig:Cu59pa} (Color online) Comparison  of $^{59}$Cu$(p,\alpha)^{56}$Ni reaction cross sections measured \cite{jsr21}, evaluated (TENDL) \cite{TENDL}, calculated by TALYS-1.95 and default options \cite{TALYS} (dotted curve), and similarly to \cite{va21} but for proton OMPs of (i-ii) Koning and Delaroche \cite{KD03} (KD) without (dash-dot-dotted) and with (short dash-dotted) DI+PE account, (iii) Saini {\it et al.} \cite{ss83} either original parameters (dash-dotted) or (iv) related $(N$-$Z)/A$ dependence (dashed), (v) the imaginary potential $W(E)$ of Ref. \cite{sk75} (solid), as well as (vi) cross--section range with the lower and upper limits given by $W$-values of 1 and 2 MeV, respectively \cite{sk79}. 
Effect of replacing $\alpha$-particle OMP \cite{va14} by \cite{va94}, for proton OMP \cite{sk75}, is also shown (short-dotted).}
\end{figure}

The same consistent parameter set has also been involved, with results for $(p,\alpha)$ reaction on $^{63,65}$Cu shown in Fig. 13 of Ref. \cite{va21}. 
Nonetheless, the calculated cross sections that are first shown as curve (i) in Fig.~\ref{Fig:Cu59pa} are obtained likewise Ref. \cite{jsr21} by using the proton OMP of Koning and Delaroche \cite{KD03}. 
They are quite close to the calculated results of the worldwide used code TALYS-1.95 \cite{TALYS} with default options including the same OMPs \cite{KD03}, the related TENDL-2019 evaluation \cite{TENDL}, and Ref. \cite{jsr21} at the center-of-mass energy of 6 MeV, while no real change corresponds to the previous minor adjustment \cite{va21} of the proton OMP \cite{KD03}. 
To understand the same factor of 2 between the measured and calculated cross sections, a summary of the rest of the presently involved model parameters is given hereafter. 

The additional nuclear--level density (NLD) parameters for the corresponding neutron--poor nuclei besides those in Ref. \cite{va21}, in the back-shifted Fermi gas (BSFG) model \cite{hv88}, are given in Table~\ref{densp}.  
For completeness of the work details, a first note may concern the larger number of the low--lying levels in an assumed complete scheme \cite{ripl3} of the target nucleus $^{59}$Cu. 
They contribute to changes below 0.2\% of the calculated $(p,\alpha)$ reaction cross section at the center-of-mass energy of 6 MeV. 
Similar changes correspond to the range of NLD parameters for the compound nucleus $^{60}$Zn \cite{va21}, with details given elsewhere \cite{va22}. 
The related $(p,\gamma)$ reaction cross section is smaller by more than two orders of magnitude. 
The level scheme above the 2.701 MeV first excited state of the residual nucleus $^{56}$Ni does not matter either. 
The feeding of even this state is only $\sim$4.4\% of the calculated $(p,\alpha)$ reaction cross section while the rest goes to the ground state (g.s.), in close agreement with experimental evidence \cite{jsr21}.

\begin{table} [b] 
\caption{\label{densp} Low-lying levels numbers $N_d$ up to excitation $E^*_d$ \protect\cite{ensdf}, used in HF calculations, and $N_d^{fit}(E^*_d)$ fitted to obtain g.s. shift $\Delta$ using average \cite{chj77} LD parameter {\it a}, and a spin cutoff factor for a variable moment of inertia \cite{va02} between half and 75\% of the rigid-body value, from g.s. to neutron separation energy, and reduced radius $r_0$=1.25 fm.} 
\vspace*{0.05in}
\begin{ruledtabular}
\begin{tabular}{ccccccc} 
Nucleus   &$N_d$&$E^*_d$& $N_d^{fit}$&$E^*_d$& $a$  & $\Delta$\hspace*{3mm} \\
          &     & (MeV) &            & (MeV) &(MeV$^{-1}$)& (MeV) \\ 
\noalign{\smallskip}\hline\noalign{\smallskip}
$^{56}$Ni &  9  & 5.353 &     9      & 5.353 & 5.5  & 2.34 \\ 
$^{59}$Cu & 38  & 3.758 &    38      & 3.758 & 6.3  &-0.23 \\
$^{59}$Zn &  5  & 1.397 &     3      & 0.894 & 6.6  &-0.75 \\
$^{60}$Zn & 12  & 3.972 &    12      & 3.972 & 6.15 & 1.00 \\ 
\end{tabular}	 
\end{ruledtabular}
\end{table}

A comment should concern the direct--interaction (DI) collective inelastic--scattering, within the distorted-wave Born approximation (DWBA) method, as well as the pre-equilibrium emission (PE) also considered \cite{va21}. 
The deformation parameters of collective states for the odd-even nucleus $^{59}$Co \cite{ck00} were used to obtain the DI proton--emission component. 
It is found to be $\sim$3.7\% of the proton reaction cross section $\sigma_R$ at the center-of-mass energy of 6 MeV while the PE similar weight was, as expected, only $\sim$0.8\%. 
Consequently, the uncommon account of DI+PE effects at so low incident energy corresponds to only $\sim$4.5\% decrease of $(p,\alpha)$ reaction cross sections shown as curve (ii) in Fig.~\ref{Fig:Cu59pa}.

On the other hand, the above--mentioned consistent parameter set was established using independently measured data as proton $\sigma_R$ and $(p,n)$ reaction cross sections \cite{va16} and validated by the analysis of $(p,\gamma)$ and even $(p,\alpha)$ reaction cross sections \cite{va21}. 
Consequently, it included a local proton OMP of Saini {\it et al.} \cite{ss83} as a better option for the Cu stable isotopes. 
However, replacing the proton global OMP \cite{KD03} by this local potential also within the analysis of $(p,\alpha)$ reaction cross section of the neutron--poor $^{59}$Cu target nucleus, we found a decrease of just $\sim$6\% connected to the curve (iii) in Fig.~\ref{Fig:Cu59pa}. 
Thus, it would correspond eventually to a standard HF cross--section range of $\sim$10\% at 6 MeV center-of-mass energy, at variance with experimental data \cite{jsr21} overestimation by a factor of $\sim$2. 

Nevertheless, a primary shortcoming of latest replacement follows the set--up of the local OMP \cite{ss83} through the fit of $(p,n)$ reaction cross sections for the stable isotope $^{65}$Cu up to an incident energy of $\sim$4 MeV \cite{ss83}, with no distinction between the isoscalar and isovector components. 
These components are duly considered in the global proton OMP between 4 and 180 MeV of Kailas {\it et al.} \cite{sk79} which was at the origin of this local potential for $^{65}$Cu. 
However only the global energy dependence of the real--potential depth $V$ was kept by Saini {\it et al.} while constant values were derived for the other local OMP parameters. 
Therefore, to adopt properly this potential for other Cu isotopes, especially off the line of stability, depth $V$=$55.5$-$0.85E$ MeV and surface--imaginary potential diffuseness $a_D$=0.57 fm  \cite{ss83} should take into account the corresponding dependencies \cite{sk79,mkm86} $V$=$50$+$24(N$-$Z)/A$+$0.4Z/A^{1/3}$-$0.85E$ MeV and $a_D$=$0.495$+$0.7(N$-$Z)/A$ fm. 
Use of the subsequent $V$ and $a_D$ values for $^{59}$Cu neutron--poor nucleus is leading to an additional $(p,\alpha)$ reaction cross--section decrease of $\sim$13\%  at 6 MeV center-of-mass energy, shown by curve (iv) in Fig.~\ref{Fig:Cu59pa}.
Although larger than the whole above--mentioned conventional HF changes, it is still insufficient to match the measured value for $^{59}$Cu target nucleus. 

On the other hand, additional attention should be given to the anomalous behavior also shown by Kailas {\it et al.} \cite{sk79,mkm86} for the surface--imaginary potential depth $W$ as a function of $A$. 
Thus, a minimum at $A$$\sim$61 has been found, followed by a steep increase within just a few mass units (Fig. 2 of \cite{sk79}). 
At the same time, the depth $W$=$3.5$-$0.3E$ MeV was found earlier by Kailas {\it et al.} \cite{sk75} by analysis of $(p,n)$ reaction on $^{59}$Co up to an incident energy of $\sim$5 MeV. 
Hence, it may be concluded that this depth should be considered rather than the constant $W$=4.1 MeV of the OMP for $^{65}$Cu \cite{ss83}.  
The corresponding results shown by curve (v) in Fig.~\ref{Fig:Cu59pa} are finally in close agreement with the measured cross section and support thus the anomalous dependence $W(A)$ \cite{sk79,mkm86} and its energy dependence for $A$$\sim$59 \cite{sk75}. 
 
The systematics in Fig. 2 of \cite{sk79} for $A$=65 indicates $W$ values between 1 and 2 MeV for target nuclei with $A$=55--59. 
The $(p,\alpha)$ reaction cross sections corresponding to these limits provide an anomalous cross--section range which has embedded the calculated excitation function using the $W(E)$ found for $^{59}$Co \cite{sk75} as well as the new experimental data for $^{59}$Cu (Fig.~\ref{Fig:Cu59pa}). 
Thus, this recent measurement supports the most pronounced sensitivity to nuclear structure effects of the imaginary--potential depth at low energies, i.e., the $W(A)$ dependence on the shell structure of the nuclei, the deformation of the target nuclei and the coupling to the collective states \cite{sk79,mkm86} altogether.
It should be noted that the proton interaction was described by the same depth $W(E)$ \cite{sk75} for the two target nuclei $^{59}$Co and $^{59}$Cu nearby the proton shell closure for $Z_T$=28. 

Nevertheless, there is a clear distinction between the changes of the standard HF calculated cross sections shown on the top of Fig.~\ref{Fig:Cu59pa} and their range associated with the anomalous proton OMP depth $W$ off the stability line. 
On the other hand, the results for $\alpha$-emission from the neutron--poor compound nucleus $^{60}$Zn are rather complementary to the also recent analysis of the $\alpha$-emission from the neutron--rich $^{59}$Mn \cite{avv21}. 
The critical role of the isovector optical potential has been entirely pointed out within this analysis too. 

For the sake of completeness, we may add that the replacement of the $\alpha$-particle OMP of Ref. \cite{va14} with an earlier, but different one \cite{va94} provides results somehow in between the conventional HF results and their range corresponding to the anomalous $W$ value in Fig.~\ref{Fig:Cu59pa}. 
Nonetheless, one must note the much simpler way to obtain the earlier OMP, particularly for $\alpha$-emission in the  mass range $A$$\sim$54, by extrapolation to low energies of an optical potential well suited at higher energies \cite{mn87}, i.e. beyond the critical OMP ambiguities. 
Therefore, it is superseded by the recent OMP \cite{va14} shown to be able to account also for $\alpha$-emission \cite{va21}. 

A similar outcome has another recent measurement also off the stability line for the excitation function of $^{54}$Fe$(p,\alpha)^{51}$Mn reaction from 9.5 to 18 MeV by Lin {\it et al.} \cite{wl22}. 
Their results have been found in agreement with the default predictions of TALYS code, including the $\alpha$-particle OMP \cite{va14}. 
Additional and complementary support for this potential has been provided by also recent  direct measurement of $^{59}$Ni$(n,p)^{59}$Co and $^{59}$Ni$(n,\alpha)^{56}$Fe reactions from 0.5 to 10 MeV, with no adjustment made to the default alpha optical potential \cite{va14} whereas the proton OMP parameters were adjusted to reproduce the low energy $(n,p)$ cross sections \cite{sak22}. 

Finally, the results of this work could be summarized as follows. 
(i) Due consideration of the proton OMP anomalies at sub-Coulomb energies for medium--weight nuclei is shown to be critical for the analysis of $^{59}$Cu$(p,\alpha)^{56}$Ni reaction. 
(ii) The 
variation in predicted cross sections from standard statistical--model calculations and the cross--section range corresponding to the anomalous proton imaginary--potential depth, for target nuclei off the line of stability, are distinct and well separated.
(iii) The new measurement of $^{59}$Cu$(p,\alpha)^{56}$Ni reaction around the energy of 6 MeV provides, under unique conditions, 
tests of proton isoscalar and isovector real--potential components, the anomalous imaginary potential \cite{sk79}, as well as previous alpha-particle OMP \cite{va14}, for nuclei off the line of stability. 
It is thus completed the similar $\alpha$-emission account by this OMP \cite{va14} for Cu stable isotopes \cite{va21} at once with all $\alpha$-induced reactions on Ni stable isotopes \cite{va16}. 

\noindent
{\it Acknowledgments.} 
The authors are grateful to the anonymous referee for detailed comments and suggestions that have helped to improve this manuscript. 
This work has been partly supported by Unitatea Executiva pentru Finantarea Invatamantului Superior, a Cercetarii, Dezvoltarii si Inovarii (Project No. PN-III-ID-PCE-2021-1260). and carried out within the framework of the EUROfusion Consortium and has received funding from the Euratom research and training programme 2014-2018 and 2019-2020 under grant agreement No 633053. The views and opinions expressed herein do not necessarily reflect those of the European Commission.

\bibliography{mybibfileA2021}

\end{document}